\documentclass[runningheads]{llncs}
\usepackage[english]{babel} 
\usepackage{microtype}
\usepackage{graphicx}
\usepackage{cite}
\begin{document}
%
\title{Automated modeling of brain bioelectric activity within the 3D Slicer environment}
\titlerunning{Automated modeling of brain bioelectric activity within 3D Slicer}
%
%
\author{Saima Safdar\inst{1} \and
Benjamin Zwick\inst{1} \and
George Bourantas\inst{1} \and
Grand Joldes\inst{1} \and
Damon Hyde\inst{2,3}
\and Simon Warfield\inst{2,3}
\and Adam Wittek\inst{1} 
\and  Karol Miller\inst{1,2}}
\authorrunning{S. Safdar et al.}
%
\institute{Intelligent Systems for Medicine Laboratory, The University of Western Australia, 35 Stirling Highway, Perth, WA, Australia \and
Harvard Medical School, Boston, MA, USA \and
Computational Radiology Laboratory, Boston Children's Hospital, Boston, MA, USA}

\maketitle              
\begin{abstract}
Electrocorticography (ECoG) or intracranial electroencephalography (iEEG) monitors electric potential directly on
the surface of the brain and can be used to inform treatment planning for epilepsy surgery when paired with numerical modeling. 
For solving the inverse problem in epilepsy seizure onset localization, accurate solution of the iEEG forward problem is critical which requires accurate representation of the patient's brain geometry and tissue electrical conductivity. 
In this study, we present an automatic framework for constructing the brain volume conductor model for solving the iEEG forward problem and
visualizing the brain bioelectric field on a 
deformed patient-specific brain model within the 3D Slicer environment.
We solve the iEEG forward problem on the predicted postoperative geometry using the finite element method (FEM) which accounts for patient-specific inhomogeneity and anisotropy of tissue conductivity.
We use an epilepsy case study to illustrate the workflow of our framework developed and integrated within 3D Slicer.

\end{abstract}


\section{Introduction}

Techniques for building patient-specific models of brain bioelectric activity and solving such models accurately and efficiently are important enablers for neuroscience and neurology. 
The application of such modeling and simulation techniques to the identification of epileptic seizure onset zones (SOZ) that consider the brain at length scales accessible by medical imaging and electrodes is of particular interest.

Epilepsy is a chronic brain disorder that causes unpredictable and recurrent electrical activity in the brain (seizures). 
It affects 65 million people worldwide, making it one of the most common neurological diseases globally \cite{thurman2011standards}.
It is estimated that about 70\% of the people living with epilepsy can be cured if diagnosed and treated properly \cite{engel_2003_greater}. 
One such treatment option is surgical resection of the area of the brain where seizures occur.
During epilepsy surgery planning using iEEG, placement of electrode grids and strips, and the body's inflammatory response to craniotomy, displace and deform the brain relative to its undeformed configuration observed in the preoperative MRI. 
This brain shift must be accurately modeled after insertion of electrode grids to enable precise localization of the SOZ with respect to preoperative MRI and postoperative CT.
The geometry of the patient's brain and the electrical conductivity distribution within the patient's head must be accurately represented for appropriate SOZ localization \cite{castano-candamil_etal_2015_solving}.

The methodology used to compute the brain shift has been extensively validated in our previous studies 
\cite{garlapati2014more, MillerRN34,miller2019biomechanical,mostayed2013biomechanical, wittek2007patient,WittekRN16,WittekRN27}.
For brain bioelectric activity modeling, continuum models based on partial differential equations are the most dominant models. 
The governing equations of the iEEG forward problem can be solved numerically using finite element (FEM) \cite{drechsler_etal_2009_full, marin_etal_1998_influence, pursiainen_etal_2011_forward, schimpf_etal_2002_dipole}, finite volume (FVM) \cite{cook_koles_2006_highresolution}, finite difference (FDM) \cite{bourantas_etal_2020_fluxconservative, hyde_etal_2012_anisotropic, saleheen_ng_1997_new, wendel_etal_2008_influence} or
boundary element methods (BEM) \cite{acar_makeig_2010_neuroelectromagnetic, meijs_etal_1989_numerical, stenroos_sarvas_2012_bioelectromagnetic}.
In our previous study \cite{zwick_etal_2022_patientspecific}, we used the FEM with a mesh coinciding with the voxel structure of the image to obtain a forward solution to the problem on a deformed patient-specific head model.

In this study, we integrate the various components of our framework in the form of independent modules 
within 3D Slicer for automatically generating the patient-specific head model for solving the iEEG forward problem. 
We demonstrate the application of our framework
by solving the iEEG forward problem for an epilepsy case study using patient data obtained from Boston Children’s Hospital. 
We solve the iEEG forward problem on the deformed 
(predicted postoperative) image obtained after using the framework of biomechanical brain modeling
\cite{saima2022computer, SafdarRN40, yu2022automatic} and compute the brain shift with methods extensively validated in our previous research
\cite{garlapati2014more, MillerRN34, miller2019biomechanical, mostayed2013biomechanical, wittek2007patient, WittekRN16, WittekRN27}.

The paper is organized as follows. 
In section 2, we describe the framework and workflow, and the methods used in each stage of the framework, for automatically constructing the patient-specific head volume conductor model for solving iEEG forward problem.
In section 3, we demonstrate the application of our framework to the solution of the iEEG forward problem.
Section 4 contains discussion
and conclusion.

\section{Methods}
\begin{figure}
    \centering
    \includegraphics[width=0.9\textwidth]{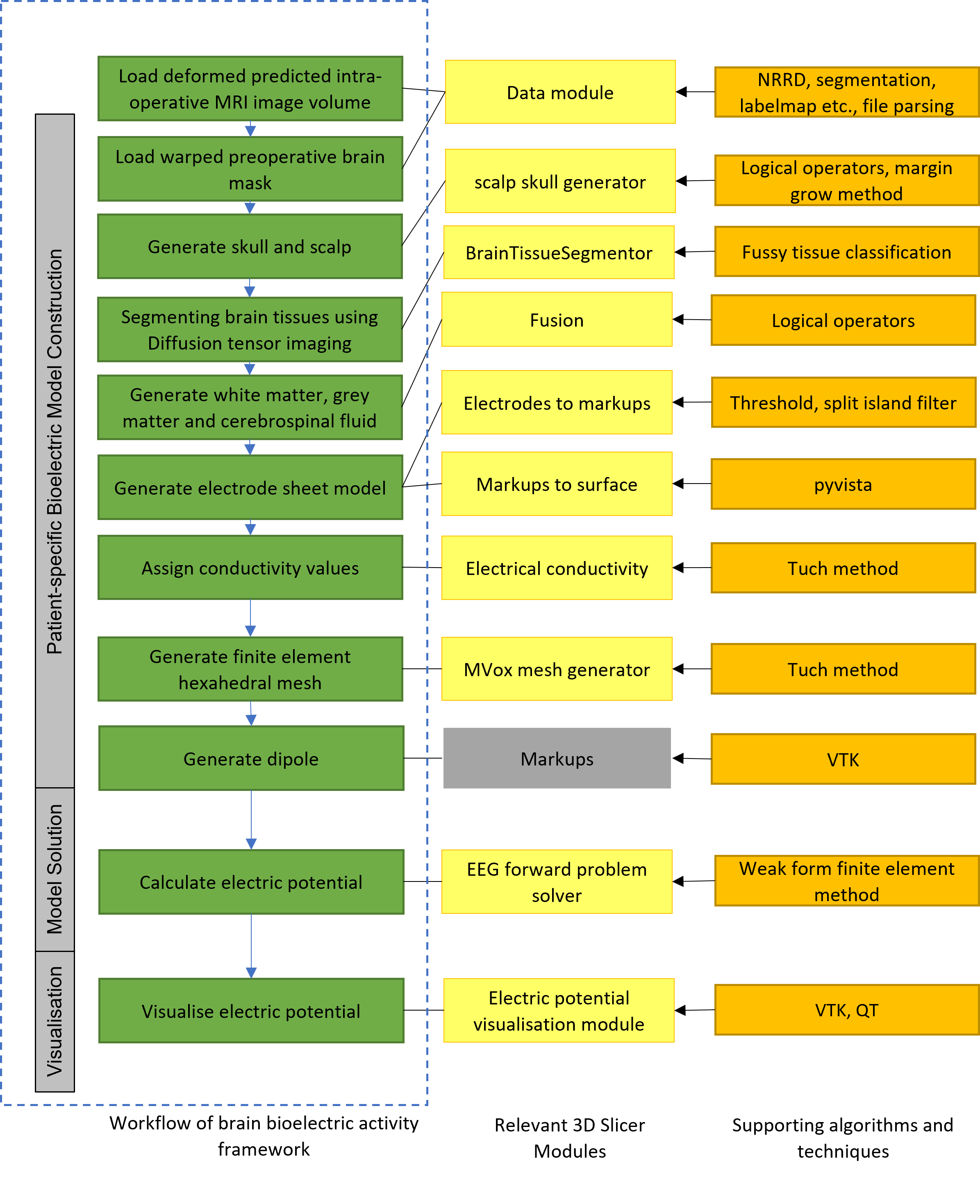}
    \caption{System workflow diagram of the automatic framework along with relevant 3D Slicer modules and supporting algorithms and techniques for patient-specific bioelectric modeling. Yellow blocks represent our developed modules and gray blocks represent the 3D Slicer modules.}
    \label{fig:workflow}
\end{figure}

The framework architecture follows a modular approach, as presented in Fig.~\ref{fig:workflow}. Our automatic system for generating a patient-specific head model and for solving the iEEG forward
problem consists of the following features:
(1) image preprocessing;
(2) constructing a patient-specific head model based on deformed predicted postoperative MRI \cite{zwick_etal_2022_patientspecific};
(3) generating electrode sheet model after extracting electrode locations from postoperative CT with electrodes implanted;
(4) assigning patient-specific electrical conductivities to brain tissues;
(5) generating a patient-specific voxel-based hexahedral mesh;
(6) defining current dipole;
(7) patient-specific model solution for iEEG forward problem; and
(8) visualization of electric potential within patient-specific head model.

The framework features are implemented as a combination of multiple 3D Slicer command-line interface (CLI) modules and scripted modules, bundled as one downloadable extension (https://github.com/SlicerCBM/SlicerCBM).  
Each module is independent, addressing only a specific requirement (see Fig.~\ref{fig:workflow}).
This modularization makes development and maintenance much easier compared to working with a monolithic code base.
Section \ref{imagePreprocessing} to section \ref{modelSolution} describe the workflow of our system for generating a patient-specific head model based on the predicted deformed postoperative MRI by our biomechanics system \cite{saima2022computer, SafdarRN40, yu2022automatic, garlapati2014more, MillerRN34, miller2019biomechanical, mostayed2013biomechanical, wittek2007patient, WittekRN16, WittekRN27} for an epilepsy case studied in our previous research, and then solving a forward problem on the generated head model using our developed 3D Slicer extensions.  

\subsection{Image preprocessing}\label{imagePreprocessing}

	


In this study, extracting the brain volume and removing the skull is performed using FreeSurfer software (http://surfer.nmr.mgh.harvard.edu/). 
It is an open source software suite for processing and analyzing human brain medical resonance images (MRIs) \cite{dale_etal_1999_cortical}. 
We used the Watershed algorithm in FreeSurfer to extract the brain portion from T1-weighted MRI \cite{SegonneRN5}. 
The Watershed algorithm uses preflooding height factor to adjust the boundaries of the brain to be included in the extracted brain volume (set of brain MRI slices without skull). 
After extracting the brain volume, we applied a threshold filter (Otsu's method) \cite{FedorovRN3} to the brain volume which selects the brain parenchyma and 
produces a brain mask. The resulting brain mask is used to reconstruct the brain geometry further in the pipeline. 

\subsection{Patient-specific head model construction for solving iEEG forward problem }

\begin{figure}
    \centering
    \includegraphics[width=\textwidth]{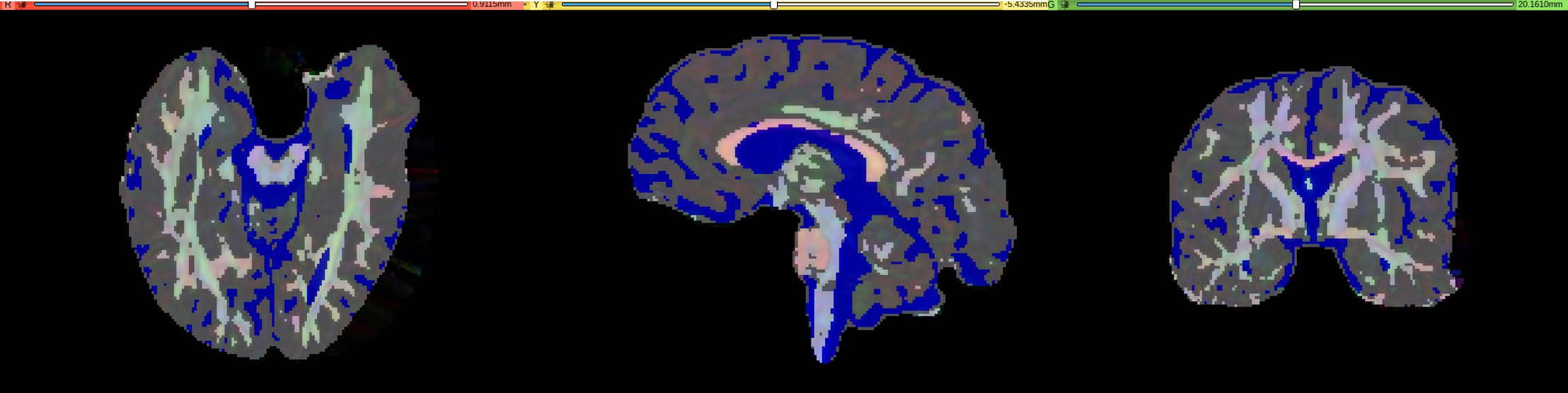}
    \caption{Results of DTI based image segmentation for a deformed DTI image. Grey represents grey matter, dark blue represents CSF and remaining represents white matter. The segmentation of the tissues (white matter, grey matter and CSF) is overlaid with the DTI image.}
    \label{fig:dti}
\end{figure}

\subsubsection{Segmenting brain tissues using diffusion tensor imaging:}
A fuzzy c-means clustering algorithm is used to classify brain tissues using diffusion tensor imaging (DTI) \cite{zwick_etal_2022_patientspecific}. In the first step, mean diffusivity is used to separate the cerebrospinal fluid (CSF) from the rest of the brain. 
In the second step, fractional anisotropy is used to 
separate the white matter (WM) from the gray matter (GM). The results of DTI based image segmentation are shown for a deformed DTI in Fig~\ref{fig:dti}.

\begin{figure}
    \centering
    \includegraphics[width=\textwidth]{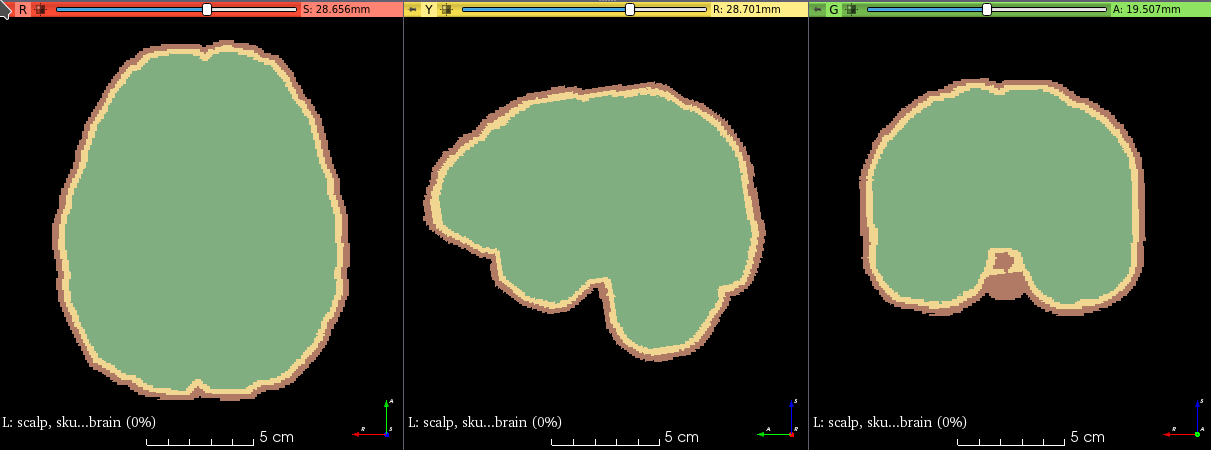}
    \caption{Scalp (light brown), skull (yellow) and brain (green) label maps within 3D Slicer}
    \label{fig:scalpSkullModel}
\end{figure}

\subsubsection{Constructing patient-specific skull and scalp segments:}
Segmenting the skull from CT is difficult due to artifacts caused by the implanted electrodes and 
requires time consuming manual segmentation,
which is not compatible with a clinical workflow. 
Therefore, instead of realistic representation of the skull geometry, 
we simplified the construction of the skull geometry by offsetting the extracted brain volume by 4 voxels. 
In a similar manner, we constructed a simplified model of the scalp by offsetting the skull model by 4 voxels. 
For constructing the simplified skull and scalp models, we developed a module “Scalp Skull Generator”, in 3D Slicer for automatically generating skull and scalp models. 
The module requires extracted brain volume, 
and two different growing margin numbers (in mm) as an input and it outputs brain, 
skull and scalp masks (mask is a separate label set of slices comprising the particular tissue type). 
The margin number defines the thickness of the skull and the scalp.
We used the following set of procedures to create the scalp, skull and brain masks: 
(1) threshold the extracted brain volume; 
(2) copy the thresholded brain volume to create a new brain volume segment;
(3) grow the new brain volume with a specified margin size (thickness in mm) to obtain the skull segment; 
(4) subtract brain volume from skull segment to create skull mask;
(5) grow the skull mask with a specified margin size (mm) to create a new segment scalp; and
(6) subtract the skull and brain from the scalp segment to get the scalp. 
We applied logical operators and margin grow methods of segment editor \cite{FedorovRN3} to construct the scalp, skull and the brain masks 
(see Fig.~\ref{fig:scalpSkullModel}).

\begin{figure}
    \centering
    \includegraphics[width=0.7\textwidth]{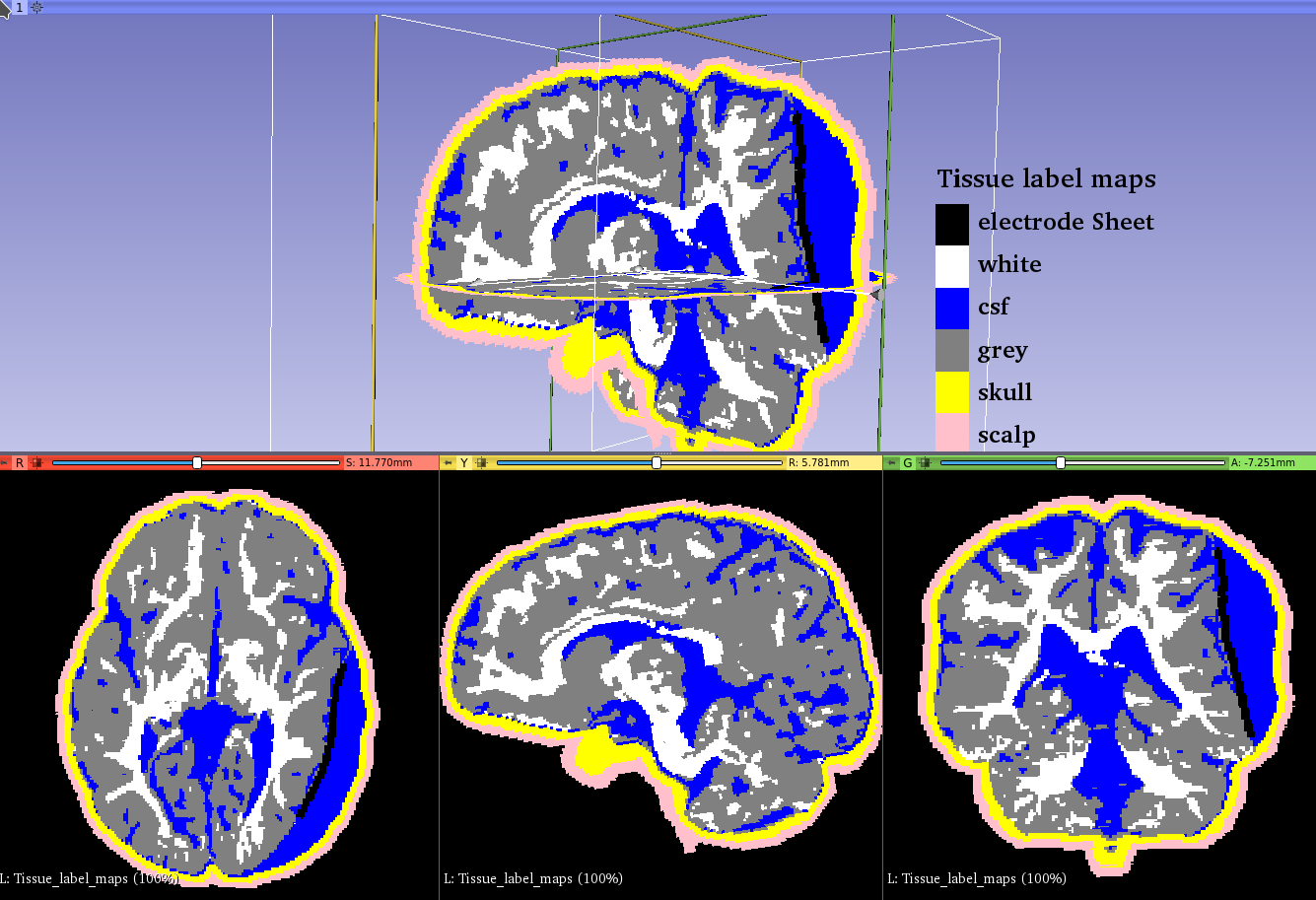}
    \caption{Tissue label maps based on deformed by insertion of electrodes postoperative image data (predicted postoperative MRI) within 3D Slicer (axial, sagittal, coronal and 3D view). Tissue classes are colored as follows:
scalp (pink); skull (yellow); GM (gray); WM (white); and CSF (blue). The location of the
electrode grid array can be identified by the line of black voxels in the vicinity of the right
temporal and parietal lobes. }
    \label{fig:tissue_label_map}
\end{figure}

\subsubsection{Combining patient-specific brain tissue segments:}
We developed the ``Fusion'' module within 3D Slicer to combine brain tissue segments (i.e., white matter, gray matter and CSF) 
with skull, scalp and electrode sheet segments. 
This module can combine a variety of tissue segments to produce the following models:
(1) undeformed model based on preoperative MRI image (with or without electrode sheet model); and
(2) deformed model based on predicted postoperative MRI (obtained by warping the preoperative MRI), with or without electrode sheet model.
We used logical operators (union) procedure of 3D Slicer segment editor module to combine all the segments.
We used the brain mask generated in 
section~\ref{imagePreprocessing} to fill any gaps in the CSF mask. 
Fig.~\ref{fig:tissue_label_map} shows tissue label maps based on deformed postoperative image data by insertion of electrodes within 3D Slicer.

\begin{figure}
    \centering
    \includegraphics[width=0.7\textwidth]{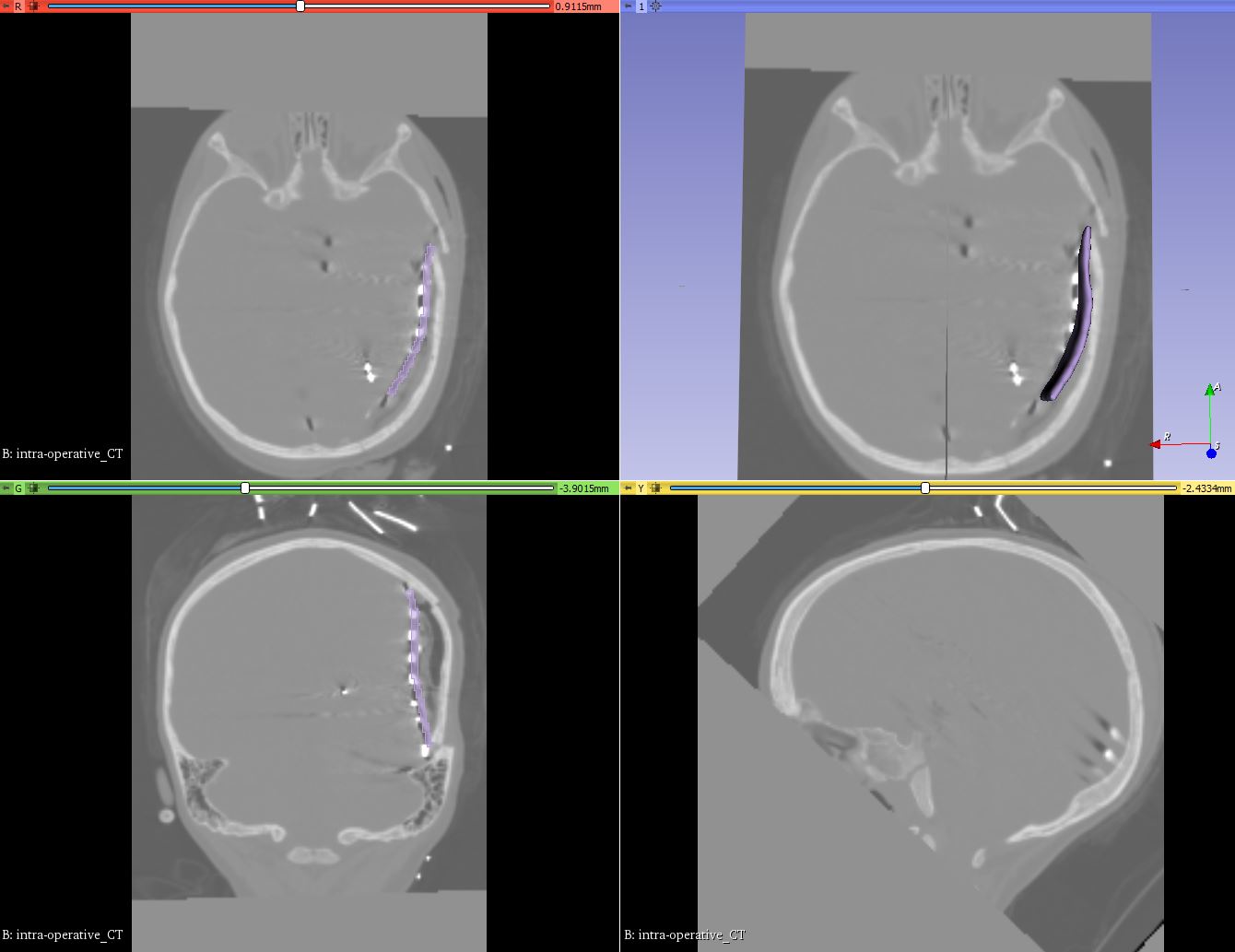}
    \caption{3D electrode sheet model overlaid with postoperative CT in axial, sagittal, coronal and 3D window within 3D Slicer}
    \label{fig:electrode_sheet}
\end{figure}

\subsubsection{Generating an electrode sheet model:}

ECoG strip and grid electrodes are typically composed of an array of platinum electrodes embedded into a sheet of a silastic material.
Our automated procedure for reconstructing the electrode sheet geometry from the postoperative CT image is as follows: 
(1) extract electrode locations from postoperative CT image; and 
(2) create the electrode sheet model.
In step 1, we extract the electrode coordinates from the
segmented electrode volume 
(image set containing electrode segmentations)
using our procedure implemented as a 3D Slicer module ``Electrodes To Markups''. 
The postoperative (post-implantation) electrode positions 
were extracted (via segmentation)
from the rigidly registered 
CT image to the preoperative MRI.
The steps involved in extracting electrode locations using 
our procedure are as follows: 
(1) create a binary label 
volume from segmented electrode volume using auto thresholding;
(2) split the 
binary label volume to individual segments corresponding to each 
electrode; and 
(3) add a point at the 
centroid of each electrode segment. 
The conversion from segmented electrode volume to binary 
label volume 
(step (1)) is performed 
using PolySeg \cite{PinterRN9}, 
a software library that provides automatic conversions between different geometry representations (e.g., label map, surface) \cite{PinterRN9}. 
Splitting the binary label volume 
(step (2)) is performed using ``split island into segments'' and then ``segment statistics'' of segment editor effect module
(step (3)) is used to get the centroids (centre of mass of the segment) \cite{PinterRN9}. 
In step 2, we used electrode locations to create 
an electrode sheet model  by means of PolyData algorithm \cite{SullivanRN39}, which creates a triangulated surface. 
We then used delaunay method to construct an electrode sheet model based on triangulated surface \cite{ValetteRN13}.

\subsubsection{Assigning patient-specific electrical conductivities:}
We assigned isotropic conductivities 
to the scalp, skull, CSF, electrode grid array substrate and gray matter regions \cite{hallez_etal_2007_review, vorwerk_etal_2014_guideline,   zwick_etal_2022_patientspecific}. We assigned anisotropic conductivity to the white matter region. We estimated the anisotropic conductivity values using the fractional methods directly from the diffusion tensors \cite{tuch_etal_2001_conductivity, zwick_etal_2022_patientspecific}.

\begin{figure}
    \centering
    \includegraphics[width=0.5\textwidth]{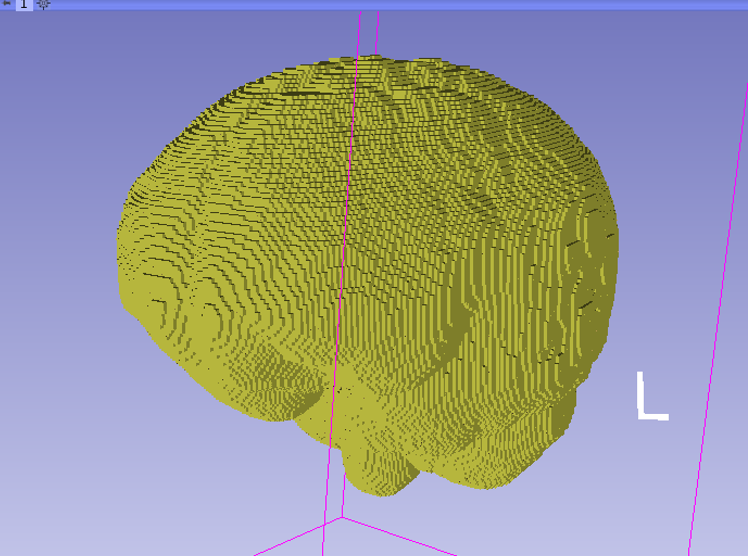}
    \caption{3D patient-specific regular hexahedral mesh in 3D window within 3D Slicer.}
    \label{fig:model}
\end{figure}

\subsubsection{Generating patient-specific voxel-based hexahedral mesh:}
We used the MVox mesh generator (https://github.com/benzwick/mvox) to generate a regular hexahedral finite element mesh (Fig. \ref{fig:model}) corresponding to the voxels of the 3D conductivity tensor image. Each nonzero voxel in the input image is converted to a hexahedral element centered at the voxel's centroid. MVox is implemented using the open source MFEM library \cite{anderson2021mfem}. It supports a variety of input and output file formats. We integrated MVox as the ``MVox Mesh Generator'' module within 3D Slicer.

\subsubsection{Defining a dipole:}
\begin{figure}
    \centering
    \includegraphics[width=0.7\textwidth]{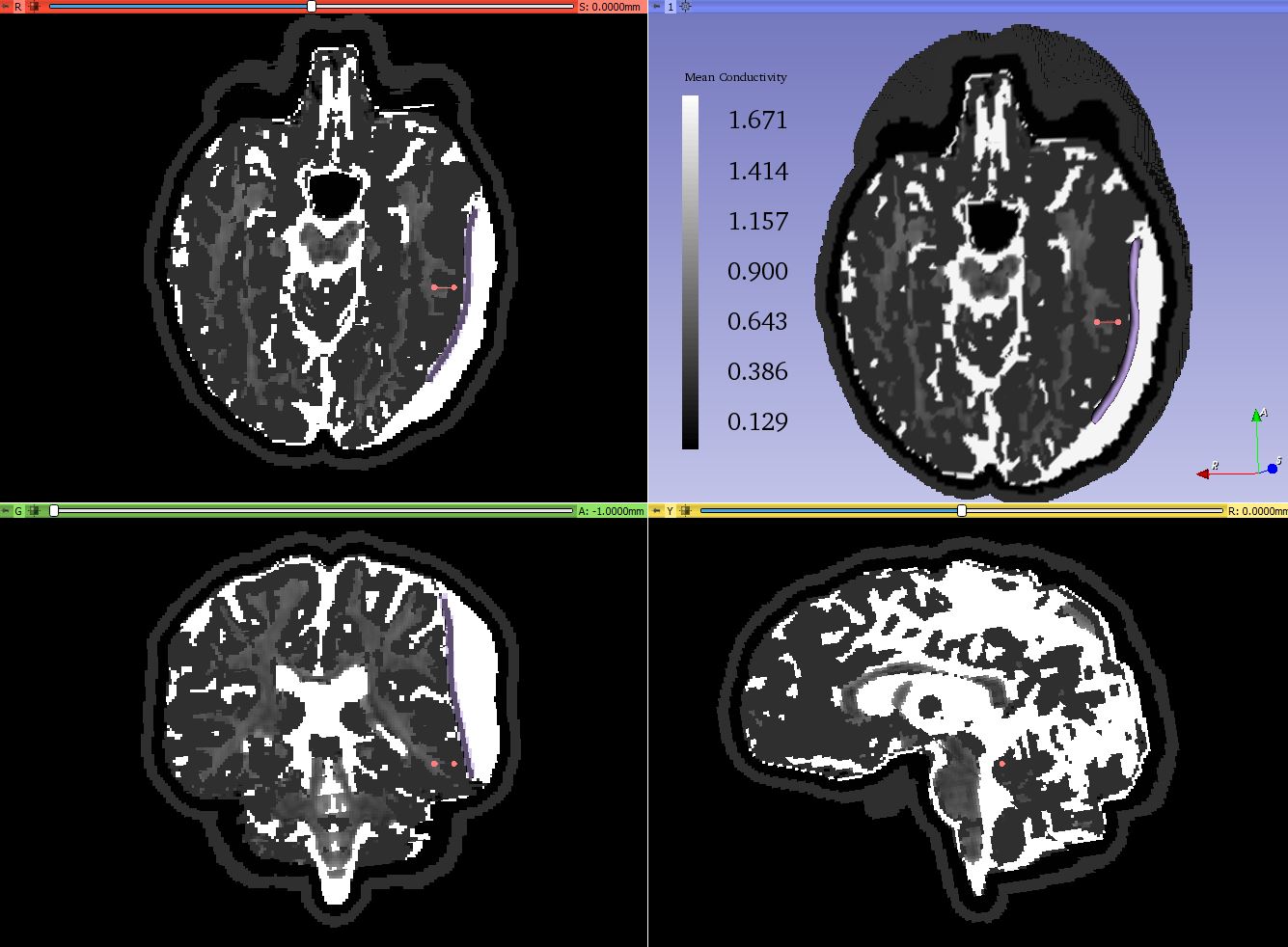}
    \caption{Mean conductivity of the deformed model based on image data deformed by implantation of electrodes shown in axial, sagittal and coronal view overlaid with deformed MRI and 3D brain model in 3D window within 3D Slicer. The current dipole moment vector is denoted by the pink line with two fiducial points. The electrode sheet is denoted by
the purple color.}
    \label{fig:mean_conduct_2}
\end{figure}

We used the ``Markups'' module of 3D Slicer \cite{FedorovRN3} to generate a vtkMRMLMarkupsLineNode to define the dipole in the gray matter region of the brain tissue. It is represented as two 3D control points in the 3D plane (see Fig. \ref{fig:mean_conduct_2})

\subsection{Model solution for EEG forward problem}\label{modelSolution}
We used the finite element method with a regular hexahedral mesh that corresponds to the voxel structure of the image \cite{schimpf_etal_2002_dipole, wolters_etal_2007_geometryadapted, rullmann_etal_2009_eeg, zwick_etal_2022_patientspecific}. We assigned conductivity tensors directly from image voxels to integration points in the elements of the hexahedral mesh. We refer to this approach of assigning the values from image voxels to elements as the ``image as a model'' approach, as the finite element mesh used to solve problem directly corresponds to image data. All elements in a regular hexahedral mesh are cubes with perfect element quality. We used a finite element mesh with 1,618,745 nodal points and 1,565,095 linear hexahedral elements for solving the EEG forward problem.
We implemented the solution to the EEG forward problem using the open-source MFEM library \cite{anderson_etal_2020_mfem} (https://mfem.org) and integrated into 3D Slicer as a separate module.
We used continuous Galerkin formulation with linear hexahedral elements for spatial discretization of the potential and for computing the electric field  \cite{zwick_etal_2022_patientspecific}. 
We used the full subtraction approach to model the current dipole source \cite{drechsler_etal_2009_full}.
We solved the discretized equations using the conjugate gradient (CG) method with an algebraic multigrid (AMG) preconditioner from the HYPRE library
of linear solvers (http://www.llnl.gov/casc/hypre). 

\begin{figure}
    \centering
    \includegraphics[width=0.5\textwidth]{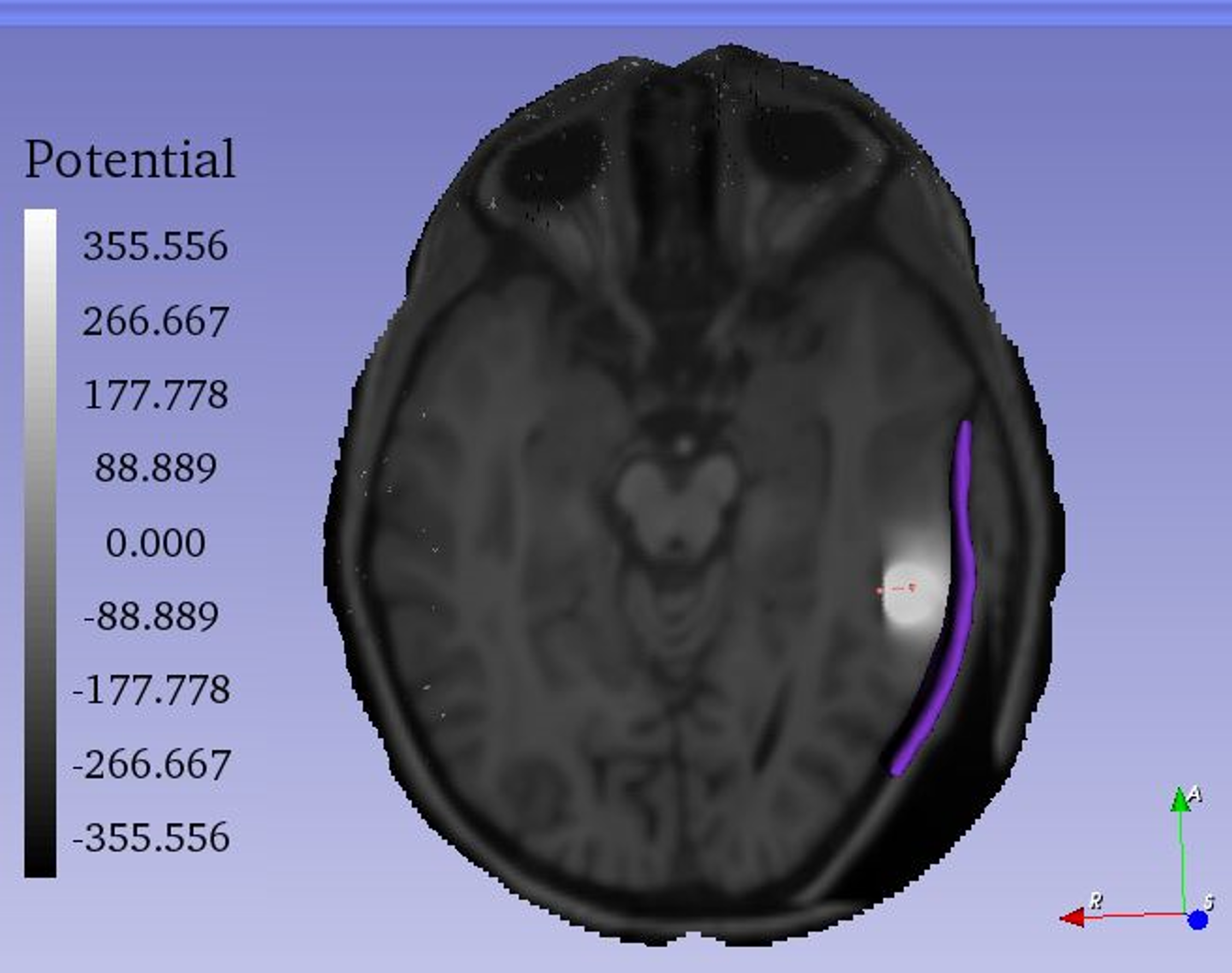}
    \caption{Electric potential in the brain generated by a current dipole predicted by the deformed model based on image data deformed by implantation of electrodes. The current dipole moment vector is denoted by the pink line with two fiducial points. The electrode sheet model is denoted by
the purple color.}
    \label{fig:electric_potential}
\end{figure}

\section{Results}
We demonstrate the application of our framework by solving an iEEG forward problem. 
We solved the iEEG forward problem with a current dipole source using the deformed image data
with the actual electrode locations. 
We put a dipole in the gray matter of the brain with a dipole moment of 100  µAmm to replicate a current dipole set up by cortical neurons.
Fig.~\ref{fig:mean_conduct_2} shows mean conductivity of the deformed model based on image data 
deformed by implantation of electrodes. Fig.~\ref{fig:electric_potential} 
shows the 
predicted distribution of electric potential within the brain computed using the model based on the deformed image data with the actual electrode positions.
We used deformed image data with actual electrode positions to avoid the error introduced by using incorrect tissue geometry. 
As discussed in our previous study \cite{zwick_etal_2022_patientspecific}, the correct geometry can reduce the effect of inaccuracy of source localization.

\section{Discussion}
In this paper, we presented an automated framework for constructing a patient-specific head volume conductor model and computing the brain bioelectric activity within the 3D Slicer environment. 
To demonstrate the application of our framework,
we used the predicted postoperative MRI with implanted intracranial electrodes obtained from our previous study \cite{zwick_etal_2022_patientspecific, saima2022computer} of a real, patient-specific case from Boston Children’s Hospital. 

The generation of the patient-specific computational
model for the iEEG forward problem which include image preprocessing, segmentation of electrodes and electrode sheet generation, scalp and skull generation, brain tissue classification and fusion of brain tissues, scalp, skull and electrode sheet, conductivity tensor estimation and generation of the finite element mesh, 
can take approximately 10 min per patient. 
The time to generate a patient-specific computational model for the iEEG forward problem using our workflow system is acceptable in the research environment and could be considered to be sufficient for clinical applications. 
We have automated all of the modeling stages in preparation for use in a clinical setting. We demonstrated the applicability of the developed framework by solving the iEEG forward problem. We showed the electric potential of the model of the brain with current dipole using the deformed (predicted postoperative) image data. 

We have developed the modules to automate the modeling tasks to reduce the time required to construct a patient-specific model for solving the iEEG forward problem
and support the medical practitioners by providing 
an easy to use graphical user interface (GUI). 
This can also make the solution of the iEEG inverse problem feasible as
we intend to combine the framework described in this 
paper with an appropriate method for 
solving the iEEG inverse problem to 
enable accurate source localization for epilepsy patients who have undergone invasive electrophysiological monitoring.
We use a modular approach to develop our system 
which makes it easy to integrate existing frameworks or components.
We plan to take advantage of this in the future
by integrating other freely available open-source frameworks (e.g.\ FreeSurfer, DUNEuro etc) within our system.

\section*{Acknowledgments}

A. Wittek and K. Miller acknowledge the support by the Australian Government through
National Health and Medical Research Project Grant no.\ APP1162030.

\bibliographystyle{splncs03_unsrt}

\bibliography{main}
\end{document}